\documentclass{svproc}
\usepackage{url}
\usepackage{graphicx}
\usepackage{multicol}
\usepackage{footmisc}
\usepackage{amssymb}
\usepackage{amsmath}
\usepackage{algorithm,algpseudocode}

\begin{document}
\mainmatter              
\title{Electrical Impedance Tomography with Box Constraint for Skull Conductivity Estimation}
\titlerunning{EIT Skull Conductivity Estimation}  
%
\author{Ville Rimpil\"ainen\inst{1} \and Theodoros Samaras\inst{2} \and Alexandra Koulouri\inst{3}}
\authorrunning{Ville Rimpil\"ainen et al.} 
%
%
\institute{University of Bath, Department of Physics, Bath, United Kingdom,\\
\email{vrimpila@gmail.com},
\and
Aristotle University of Thessaloniki, Department of Physics, Thessaloniki, Greece \\
\and Tampere University, Faculty of Information Technology and
Communication Sciences, Tampere, Finland}

\maketitle              

\begin{abstract}
Unknown electric conductivities of human tissues is a common issue in medical engineering. Electrical impedance tomography (EIT) is an imaging modality that can be used to determine these conductivities {\it in vivo} from boundary measurements. In this paper, we demonstrate that local conductivity values of different skull segments can be solved from EIT measurements with the help of a box constraint. Based on our numerical results, the accuracy of the results depended on the locations of the current carrying electrodes and the signal to noise ratio of the measurements. Particularly, the conductivity values of the skull segments that located below the current carrying electrodes were reconstructed more accurately.
%
\end{abstract}
%

\section{Introduction}

The determination of the electric conductivity values of the different tissues of the human head is an important problem that is often encountered in electroencephalography source imaging and transcranial direct current stimulation. The efficacy of these modalities is highly dependent on the knowledge of the tissue conductivities, particularly the highly insulating skull \cite{mon14,sch15}.

Electrical impedance tomography (EIT) is an imaging modality that can be used to solve the internal conductivity profiles with the help of boundary electrodes \cite{hol04}. Most EIT studies of the human head are limited to only very few (4 or less) tissue conductivity values \cite{dab16,fer18} which effectively neglects any possibility to find local variations within any of the tissue compartment. Previously, it has been reported that, for example, the electric conductivity of the temporal bones can be much lower than elsewhere in the skull \cite{law93}.

In the current study, we demonstrate reconstruction of local conductivity variations in the skull compartment by using a box constraint. We have divided the skull in pieces in order to reconstruct different conductivity values of each segment. This work is a simulation study that is carried out by using 3-dimensional finite element (FE) -based head models.

\section{Theory}

\subsection{Observation model}

The observation model of EIT can be written as
\begin{equation}\label{eq:Observation}
V_\mathrm{meas} =U (\sigma,I) + v_{\mathrm{noise}},
\end{equation}
where $V_\mathrm{meas}\in\mathbb{R}^M$ are the measured boundary
voltages, $U(\sigma,I)$ is the mapping that connects the
conductivity distribution $\sigma$ and the pre-defined current
injections $I$ with the measured voltages, and $v_{\mathrm{noise}}$
represents the measurement noise. For the numerical solution of the
EIT problem, the conductivity distribution is discretized and
approximated as $\sigma(x) = \mathrm{\Sigma}_{i=1}^n \sigma_i
\phi_i(x)$, where $\phi_i(x)$ can be linear basis
functions, for example, and $n$ is the total number of nodes of the FE head model \cite{vau99}.

The mapping $U(\sigma,I)$ can be linearized with respect to the
conductivity parameterization by using the Taylor series expansion
\begin{equation}\label{eq:taylor}
U(\sigma) = U(\sigma_{\mathrm{lin}})+J(\sigma-\sigma_{\mathrm{lin}})+
\mathcal{O}(\|(\sigma-\sigma_{\mathrm{lin}})\|^2),
\end{equation}
where $\sigma\in\mathbb{R}^n$, $J\in\mathbb{R}^{M\times n}$,
$J_{j,i}=\frac{\partial U_j(\sigma_{\mathrm{lin}})}{\partial
\sigma_i}$ is the ($j,i$)th element of the Jacobian matrix \cite{vau97}
evaluated at a given linearization point
$\sigma_{\mathrm{lin}}\in\mathbb{R}^n$.

From (\ref{eq:Observation}) and (\ref{eq:taylor}), we can approximate the observation model as
\begin{equation}\label{eq:Observation2}
V_\mathrm{meas}= U(\sigma_{\mathrm{lin}})+J(\sigma-\sigma_{\mathrm{lin}})+ v_{\mathrm{noise}}.
\end{equation}

\subsection{EIT inverse problem with box constraint}
\label{boxxx}

Here, the unknown conductivity parameters $\sigma$ of
(\ref{eq:Observation2}) are solved with the help of a box
constraint. The box constraint can be used to define the lower and
upper bounds for the unknown conductivities ($\sigma_L$ and
$\sigma_U$, respectively), and the corresponding minimization
functional is of the form
\begin{equation}\label{eq:OriginalProblem}
\underset{\sigma}{\;\mathrm{min}}\,f(\sigma)=\|b-J\sigma\|_2^2+\lambda\|B\sigma\|_2^2
\,\mbox{ subject to $\sigma_L\leq\sigma\leq
 \sigma_U$},
\end{equation}
where $b = V_\mathrm{meas}-U(\sigma_{\mathrm{lin}})+J\sigma_{\mathrm{lin}}$, $\lambda$ is a
regularization coefficient and $B$ represents the conductivity prior.
The problem is now solved iteratively by using algorithm \ref{alg:EIT_Iter}. 

\begin{algorithm}[htb]
\caption{Solve the EIT imaging problem} \label{alg:EIT_Iter}
    \begin{algorithmic}[1]
    \State \textbf{Initialization:} $\sigma^{(0)}=0.5(\sigma_U-\sigma_L)+\sigma_L$ and $U^{(0)}=U(\sigma^{(0)},I)$;
    \For{$i=0,\ldots,N_{\mathrm{max}}$}
               \State Update Jacobian  $J^{(i)}$ and $b^{(i)} = V_\mathrm{meas}-U^{(i)}+J^{(i)}\sigma^{(i)}$;
               \State $\sigma^{(i+1)}:=\min_{\sigma}\|b^{(i)}-J^{(i)}\sigma\|_2^2+\lambda\|B\sigma\|_2^2$ subject to $\sigma_L\leq\sigma\leq \sigma_U$;
                \State Estimate 
                $U^{(i+1)}=U(\sigma^{(i+1)},I)$;
                \If {$\|V_\mathrm{meas}-U^{(i+1)}\|_2\leq \|v_{\mathrm{noise}}\|_2$ and $\|\sigma^{(i+1)}-\sigma^{(i)}\|_2<\varepsilon_{\mathrm{tol}}$} {Quiet} \EndIf
            \EndFor
    \end{algorithmic}
\end{algorithm}

\section{Methods}

A numerical head model for the human head was downloaded\footnote{\url{http://eidors3d.sourceforge.net/data_contrib/at-head-mesh/at-head-mesh.shtml}} \cite{tiz07}. It consisted of 262 368 elements joined in 48 394 nodes, and 33 electrodes that were place on the scalp, see Figure 1. In the head model, the scalp, skull, cerebro-spinal-fluid (CSF) and brain compartments were segmented. For this study, the skull compartment was further divided into $c=5$ segments (see, Figure 1) which approximately corresponded the frontal, parietal, occipital and left and right temporal bone. The conductivity  value of each segment was subsequently estimated by using EIT and a box constraint. The remaining conductivity parameters were considered as known.

\begin{figure}\label{fig1}
\includegraphics[width=10cm]{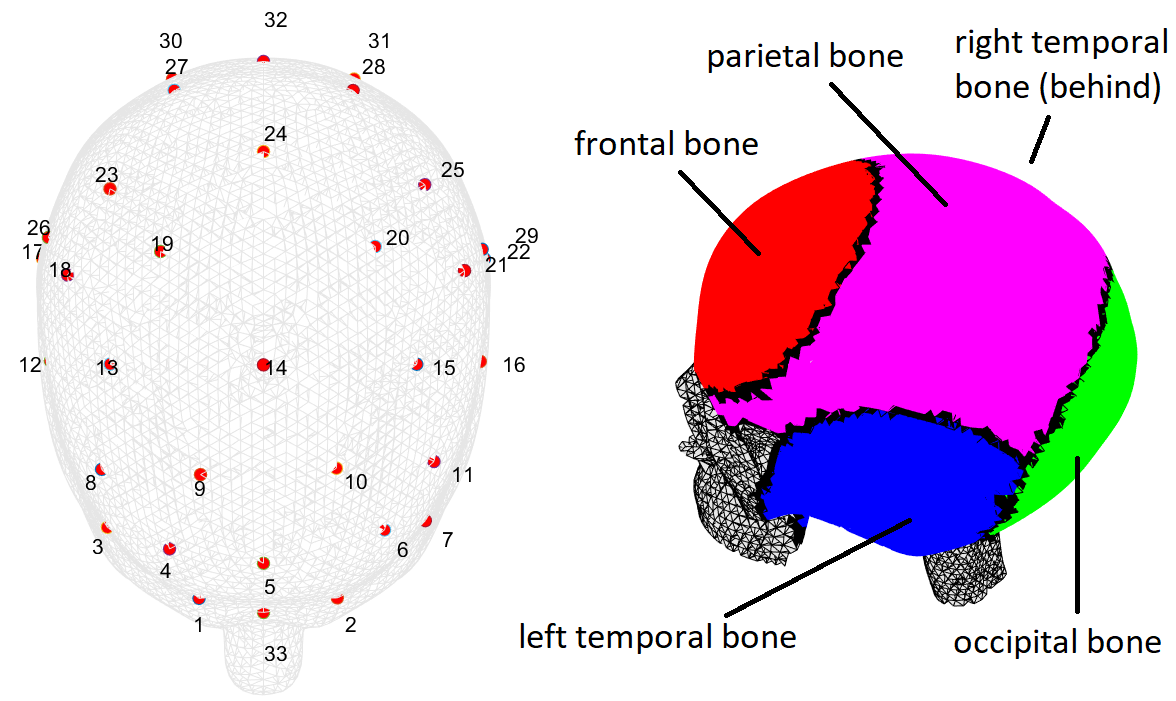}
\caption{Left: The numerical head model with the scalp electrodes that were used in the simulations. Right: The skull was roughly segmented to frontal, parietal, occipital and left and right temporal bone. The conductivity values of these segments were reconstructed with the help of EIT and box constraint.}
\end{figure}

\subsection{Jacobian matrix for skull segments}

For the computations, the Jacobian matrix $J\in\mathbb{R}^{M\times n}$ needed to be re-defined for the $c$ 
skull segments (i.e. $J\in\mathbb{R}^{M\times c}$). First, the columns of the Jacobian matrix that corresponded to known
conductivities were discarded reducing the size of the Jacobian matrix to ${M\times n_\mathrm{sk}}$ where $ n_\mathrm{sk}$ is
the number of the FE nodes in the skull compartment. Then, by defining a 1-to-1 projection 
$P\in \mathbb{R}^{n_\mathrm{sk}\times{c}}$ between the different
segments and the skull nodes, the size of the Jacobian was
further reduced to ${M\times c}$. 

\subsection{Implementation of box constraint}
\label{boximple}

The problem (\ref{eq:OriginalProblem}) was
solved by using the interior point method (IPM) \cite{Boyd2004}. For
this purpose, the problem (\ref{eq:OriginalProblem})
was first reformulated as an unconstrained problem with the help of the logarithmic
barrier method \cite{Boyd2004}
\begin{equation}\label{eq:MinimizationWithBarrier}
\underset{\sigma}{\;\,\mathrm{min}} \, \Phi_t(\sigma)=
t\left(\|b-J\sigma\|_2^2+\lambda\|B\sigma\|_2^2\right)-\sum_{i=1}^c\log{[
(\sigma_i-\sigma_{L,i})(\sigma_{U,i}-\sigma_i)]},
\end{equation}
where $\sigma\in\mathbb{R}^c$ is the unknown conductivity vector, $\sigma_{L,i}$ and $\sigma_{U,i}$ are the lower and upper
conductivity limits in each segment.

The solution is then achieved by solving a sequence of
unconstrained problems (\ref{eq:MinimizationWithBarrier}) for
increasing value of $t>0$.
The IPM algorithm consists of an inner and outer loop \cite{Boyd2004}. In the
inner loop, problem (\ref{eq:MinimizationWithBarrier}) is solved
by keeping the value of $t> 0$ constant while in the outer loop, the value of $t$ is
updated. The IPM algorithm
terminates based on a chosen stopping criterion. Here, we used the
duality gap as in \cite{Kim2007,Koh2007}.

\subsubsection{Stopping Criterion} 

Duality gap is the difference between the value of $f(\sigma)$
(\ref{eq:OriginalProblem}) and its dual functional $g$
given by
\begin{equation}\label{eq:dualitygap}
\eta=f({\sigma})-{g}({p},{q},{\mu},{\nu}),\end{equation} where
${g}(p,q,{\mu,\nu})
=-0.25\left(p^{\mathrm{T}}p+\lambda{q^\mathrm{T}q}\right)-p^{\mathrm{T}}b+\mu^{\mathrm{T}}\sigma_L-\nu^\mathrm{T}\sigma_U$.

Given a solution ${\sigma}$ for problem (\ref{eq:OriginalProblem}), we estimate $g$ 
based on
\begin{equation}\label{eq:dualVariables}
{p}= 2(J{\sigma}-b),\,\,\,{q}=2
B{\sigma},\,\,\,\mu=\max(0,[J^\mathrm{T}{p}+B^\mathrm{T}{q}])\,\,\mbox{and}\,\,\nu=\max(0,-[J^\mathrm{T}{p}+B^\mathrm{T}{q}]).
\end{equation}

\subsubsection{Inner loop} 

Keeping  $t$ fixed, the problem (\ref{eq:MinimizationWithBarrier}) can be solved with the help of the Newton
method \cite{Heath2002} followed by backtracking line search
(BLS) \cite{Boyd2004}. In particular, we estimated first the search
direction defined as $\Delta \sigma\in \mathbb{R}^{c}$
by solving the linear system
\begin{equation}\label{eq:NewtonSystem}
H_{\Phi_t}\Delta \sigma=-G_{\Phi_t},
\end{equation}
where $H_{\Phi_t}= \nabla^2\Phi_t(\sigma)\in\mathbb{R}^{c\times c}$
is the Hessian matrix and $G_{\Phi_t}= \nabla
\Phi_t(\sigma)\in\mathbb{R}^{c}$ is the gradient of $\Phi_t(\sigma)$,
respectively. Then the updated
solution is ${\sigma}:=\sigma+s\Delta\sigma$ where
$s > 0$ is the step size estimated by the BLS.

\subsubsection{Outer loop}

The outer loop of the IPM updates the value of $t$
\cite{Koh2007,Kim2007}. 
Here, we used
\begin{equation}\label{eq:tUpdateRule}
t:=\left\{ \begin{array}{ll}
 \max\{\alpha\min\{c/\eta,t\},t\}, & s\geq s_\mathrm{min}\\
  t &\ s<s_\mathrm{min},
       \end{array} \right.
\end{equation}
where $s$ is the estimated step size and $s_\mathrm{min}=0.7$
is the threshold that determines whether $t$ will be updated or not.
This $t$-rule ensures that $t$ stays constant until the cost
function $\Phi_t$ (\ref{eq:MinimizationWithBarrier}) is nearly
minimized, i.e. $\|\nabla\Phi_t\|_2\simeq 0$.

\subsection{Simulation set-up}

In the numerical experiments, two EIT current injections with
amplitude 2 mA were used: the electrodes 21 and 18 were used in the
first, and 6 and 24 in the second current injection. The resulting voltages on the
electrodes, excluding current carrying electrodes, were measured
against a common ground that was placed on the chin. Random Gaussian
white noise was added in the measurements. 

In the test cases, we used the following tissue conductivity values (in S/m): $\sigma_{\mathrm{scalp}} = 0.43$ \cite{ram04}, $\sigma_{\mathrm{CSF}} = 1.79$ \cite{bau97} and $\sigma_{\mathrm{brain}} = 0.33$ \cite{ram04}. The skull conductivity values were varied around on the literature value 0.01 S/m that was found in \cite{dan11}. The tested conductivity values of the frontal, parietal, occipital and left and right temporal bone segment are given in Table 1. The rest of the skull was kept in a fixed (known) conductivity value. 

The EIT inverse problem with the box constraint was solved as detailed in Section \ref{boximple} in order to estimate the unknown skull conductivity parameters. The lower and upper limits of the box constraint were $\sigma_L=0.001$ S/m and $\sigma_U=0.04$ S/m, respectively.
Since the number of measurements
$M=62$ exceeded the number of unknown skull conductivity parameters $c=5$, thus the problem 
(\ref{eq:OriginalProblem}) was over-determined, we could set the
regularization parameter $\lambda=0$ and effectively omit the prior term. 

\section{Results and discussion}

The conductivity values of the 5 skull segments and the signal to noise ratios (SNR) used in the test cases, and the reconstructed conductivity values are summarized in Table 1. The reported conductivity values are averages of 10 reconstructions (computed with different noise realizations), and the uncertainty values are the standard deviations of these 10 values.

\begin{table}
\caption{The true and reconstructed conductivity values (in $10^{-3}$ S/m) of the different skull segments.}
\begin{center}
\begin{tabular}{ c|c|c|c|c|c }
     SNR          & Frontal                 & Occipital               & Right temporal      & Left temporal      & Parietal \\
\hline
True              & 10                       & 10                        & 10                        & 10                      & 10  \\
40 dB &  10.0 $\pm$ 0.0 &  9.7 $\pm$ 0.5 &  10.4 $\pm$ 0.3 &  10.1 $\pm$ 0.3 &  10.1 $\pm$ 0.1 \\ 
20 dB &  9.7 $\pm$ 0.4 &  9.6 $\pm$ 2.8 &  6.5 $\pm$ 2.7 &  12.6 $\pm$ 2.4 &  9.9 $\pm$ 0.3 \\
 \hline
True              & 10                       & 10                        & 7                          & 7                        & 10  \\
40 dB &  10.0 $\pm$ 0.1 &  10.1 $\pm$ 0.3 &  7.4 $\pm$ 0.3 &  6.7 $\pm$ 0.2 &  9.9 $\pm$ 0.1 \\
20 dB &  9.5 $\pm$ 0.3 &  5.4 $\pm$ 3.1 &  2.6 $\pm$ 1.6 &  16.8 $\pm$ 4.3 &  10.8 $\pm$ 0.7 \\
 \hline
True              & 10                       & 10                        & 5                          & 5                        & 10  \\
40 dB &  9.9 $\pm$ 0.1 &  9.0 $\pm$ 0.7 &  4.6 $\pm$ 0.3 &  4.7 $\pm$ 0.2 &  10.0 $\pm$ 0.0 \\
20 dB &  10.3 $\pm$ 0.2 &  10.7 $\pm$ 2.8 &  3.3 $\pm$ 2.1 &  4.6 $\pm$ 2.2 &  10.5 $\pm$ 0.5\\
 \hline
True              & 10                       & 10                        & 1                          & 1                        & 10  \\
40 dB &  10.1 $\pm$ 0.0 &  10.9 $\pm$ 0.7 &  1.0 $\pm$ 0.0 &  1.0 $\pm$ 0.0 &  10.0 $\pm$ 0.0\\
20 dB &  10.2 $\pm$ 0.5 &  11.7 $\pm$ 3.7 &  4.6 $\pm$ 3.5 &  3.9 $\pm$ 1.8 &  10.4 $\pm$ 0.6\\
 \hline
True              & 7                         & 7                          & 5                          & 5                        & 7  \\
40 dB &  6.9 $\pm$ 0.1 &  7.4 $\pm$ 0.2 &  5.1 $\pm$ 0.1 &  5.0 $\pm$ 0.1 &  7.0 $\pm$ 0.0 \\
20 dB &  7.0 $\pm$ 0.3 &  9.4 $\pm$ 1.5 &  13.3 $\pm$ 5.3 &  4.8 $\pm$ 1.6 &  7.0 $\pm$ 0.3\\
 \hline
True              & 7                         &  7                         & 1                          & 1                        & 7  \\
40 dB &  7.0 $\pm$ 0.1 &  7.3 $\pm$ 0.3 &  1.2 $\pm$ 0.2 &  1.5 $\pm$ 0.3 &  6.9 $\pm$ 0.0 \\
20 dB &  6.0 $\pm$ 0.5 &  15.1 $\pm$ 6.3 &  5.6 $\pm$ 4.9 &  1.3 $\pm$ 0.4 &  6.4 $\pm$ 0.5\\
 \hline
True              & 7                         &  9                         & 10                          & 10                        & 8  \\
40 dB &  7.1 $\pm$ 0.1 &  8.8 $\pm$ 0.2 &  9.9 $\pm$ 0.3 &  10.0 $\pm$ 0.2 &  8.0 $\pm$ 0.0\\
20 dB &  7.1 $\pm$ 0.3 &  14.5 $\pm$ 1.8 &  15.9 $\pm$ 1.5 &  24.1 $\pm$ 6.4 &  6.9 $\pm$ 0.5 \\
 \hline
True              & 10                         &  11                         & 13                        & 14                        & 9  \\
40 dB &  10.0 $\pm$ 0.1 &  11.4 $\pm$ 0.5 &  13.2 $\pm$ 0.3 &  14.3 $\pm$ 0.3 &  9.0 $\pm$ 0.0 \\
20 dB &  10.7 $\pm$ 0.2 &  7.1 $\pm$ 1.7 &  22.8 $\pm$ 4.7 &  18.7 $\pm$ 4.0 &  9.0 $\pm$ 0.2 \\
 \hline
\end{tabular}
\end{center}
\end{table}

It can be seen that in all the 40 dB test cases the reconstructed conductivity values match well with the true ones, and that the accuracy of the results deteriorates with increasing noise, as expected. The uncertainty values depend strongly on the location: the standard deviation values for occipital and both temporal bones are higher than for frontal and parietal bones, particularly this is evident for the 20 dB test cases. This is due to the choice of the current carrying electrodes which all located above the occipital and temporal bones. This implies that better accuracy for the occipital and temporal bones could be achieved by using current injections through electrodes that locate above these areas. 

In addition, in cases in which the bone segments had conductivity values equal to the lower limit of the box constraint, a value still within the box that was very close to the box limit was found. This verifies that the box constraint algorithm worked correctly.

\section{Conclusions and future work}

In this paper, we demonstrated the use of a box constraint when reconstructing conductivity values from EIT measurements for different segments of the human skull. The accuracy of the results depended on the SNR and whether current carrying electrodes located above the unknown skull segment. In the future, EIT studies with further refinement in the parameterization of the skull conductivity (i.e. smaller segments) and additional unknown tissue compartments will be tested.

\section*{Acknowledgements}

The authors wish to thank Dr. Nathan D. Smith for his insight on the dual function. This project has received funding from the ATTRACT project funded by the EC under Grant Agreement 777222 and from the Academy of Finland post-doctoral program (project no. 316542).

\section*{Conflict of interest}

The authors declare that they have no conflicts of interest.

\bibliographystyle{spphys}

\end{document}